\def\year{2022}\relax
\documentclass[letterpaper]{article} 
\usepackage{aaai22}  
\usepackage{times}  
\usepackage{helvet}  
\usepackage{courier}  
\usepackage[hyphens]{url}  
\usepackage{graphicx} 
\usepackage{natbib}  
\usepackage{caption} 
\DeclareCaptionStyle{ruled}{labelfont=normalfont,labelsep=colon,strut=off} 
\usepackage{algorithm}
\usepackage{algorithmic}
\usepackage{xcolor}

\usepackage{newfloat}
\usepackage{listings}
\floatstyle{ruled}
\newfloat{listing}{tb}{lst}{}
\floatname{listing}{Listing}
\usepackage{booktabs}

\title{Analyzing the Engagement of Social Relationships During Life Event Shocks in
Social Media}
\author {
    Minje Choi,\textsuperscript{\rm 1}
    David Jurgens, \textsuperscript{\rm 1}\textsuperscript{,}\textsuperscript{\rm 2}
    Daniel M. Romero \textsuperscript{\rm 1}\textsuperscript{,}\textsuperscript{\rm 2}\textsuperscript{,}\textsuperscript{\rm 3}
}
\affiliations {
    \textsuperscript{\rm 1} School of Information, University of Michigan\\
    \textsuperscript{\rm 2} Computer Science and Engineering Division, University of Michigan\\
    \textsuperscript{\rm 3} Center for the Study of Complex Systems, University of Michigan\\
    \{minje, jurgens, drom\}@umich.edu
}

\begin{document}

\maketitle

\begin{abstract}
Individuals experiencing unexpected distressing events, \textit{shocks}, often rely on their social network for support. While prior work has shown how social networks respond to shocks, these studies usually treat all ties equally, despite differences in the support provided by different social relationships. Here, we conduct a computational analysis on Twitter that examines how responses to online shocks differ by the relationship type of a user dyad. We introduce a new dataset of over 13K instances of individuals' self-reporting shock events on Twitter and construct networks of relationship-labeled dyadic interactions around these events. By examining behaviors across 110K replies to shocked users in a pseudo-causal analysis, we demonstrate relationship-specific patterns in response levels and topic shifts. We also show that while well-established social dimensions of closeness such as tie strength and structural embeddedness contribute to shock responsiveness, the degree of impact is highly dependent on relationship and shock types. Our findings indicate that social relationships contain highly distinctive characteristics in network interactions and that relationship-specific behaviors in online shock responses are unique from those of offline settings.
\end{abstract}

\section{Introduction}
\label{sec:intro}
Not everything in life comes with preparation: People may experience sudden events such as the death of a loved one or a sudden job loss. Exposure to such events can cause adverse effects on one's mental~\cite{burton2006bereavement} and financial status ~\cite{atkinson1986social}. To mitigate such harmful effects that arise from experiencing unexpected life events or \textit{shocks}, people often ask for help through their own accessible social network, which is made of ties  belonging to different categories or relationships such as friends, family, or workplace relationships. The different relationships are known to provide varying levels and types of support, which contribute to the overall support network of an individual~\cite{vaux1985support}. 
Here, we study behavioral differences caused by relationship types by comparing responses to shock events.

In online settings, shocks like a sudden change in a user's status can cause behavioral changes from other users in response~\cite{oh2016impression}. Prior studies have shown online shocks are associated with a variety of network-level reactions such as changes in centralization levels~\cite{zhang2017shocking} and  increased communication among closer ties~\cite{hobbs2017connective,romero2016stress}. Furthermore, users experiencing shock events may choose to publicly disclose it on social media, a behavior which has been increasingly studied by social researchers~\cite{haimson2018breakup,andalibi2019disclosing}. However, these studies have largely assumed that all ties are equal rather than modeling the potential interaction between different relationships and shocks, and little is yet known of the relationship-specific response behaviors towards these disclosures. We pursue this open question, asking whether relationship-specific shock behavior in online platforms mirrors that of the offline world, or whether structural and normative constraints of online social networks may cause individuals to interact differently with their relationships when compared to offline settings.

In this study, we conduct a large computational analysis of responses to shocks in online social networks to test how interpersonal relationships engage. We introduce a new dataset of over 13,000 Twitter users who posted shock events along with their interactions with others, each labeled with their inferred relationships to the shocked user~(Section \ref{sec:data}). 
Using causal inference methods, we approximate the effect of experiencing and posting shock events on receiving responses from Twitter users, and how these activation levels differ both in magnitude and significance depending on both the type of relationship and the type of shock~(Section~\ref{sec:diff}). To understand the interaction between shock event types, relationships and responding behavior, we analyze how tie strength and structural embeddedness influence the users in different relationships to reply to a shock, both strongly recognized network properties for determining interactions in social networks~(Section~\ref{sec:closeness}). Finally, we identify relationship-specific differences in the content of shock responses by measuring topic shift via a topic model~(Section~\ref{sec:topic}).

Our contributions are as follows. First, we demonstrate a method for identifying and extracting instances of shock experiences from Twitter posts using active learning and result in a corresponding dataset of $\sim$13K users experiencing shocks and 179,563 users making a total of 110,540 replies to them. This data is augmented by labeling the relationship between shocked and replying users and adding a corresponding control set of users, matched for aspects such as demographics, location, and activity level.
Second, through a large-scale quasi-causal analysis, we demonstrate how relationship types determine levels of responsiveness and topic shift in the responses to shock tweets that are posted online. Our findings partially align with existing theories on social relationships in offline settings; however, we also discover contrasting results that may relate to differences between online and offline settings. For instance, we observe that romance and family relationships generally respond less to shocks than social relationships, which may be due to the existence of other communication channels preferred over Twitter's public space. 
Third, we show that tie strength and network embeddedness each have different effects in predicting responsiveness specific to the relationship type and shock. These results point to the existence of unique social dynamics for each relationship type and suggest how individuals and their supporters can better mobilize their social networks in times of unexpected distress.

\section{Shocks, relationships, and networks}
\label{sec:litreview}
Separate strands of research have examined shocks and social relationships in times of stress, creating expectations for different social ties that might interact in these events. Here, we outline the major work in each area to motivate specific research questions pursued in our study.

\subsection{Shocks and Engagement in Social Networks}
A shock is defined as an unforeseen event capable of disrupting an individual,  a group, or a social network~\cite{jackson1988discerning}. Prior studies have looked at how exposure to exogenous shocks such as community censorship~\cite{zhang2017shocking}, sudden price changes~\cite{romero2016stress} or disasters~\cite{corbo2016exogenous} affected the network's communication behavior such as contact frequency and clustering tendency.  

A particular category of shocks widely studied across various academic fields is that of individuals experiencing unexpected life events during their life courses and how these events affect their social networks. Events such as the death of a family member or unexpected pregnancy can increase intra-family strain and harm well-being levels ~\cite{lavee1987effect}, which can also develop into health or depression issues~\cite{kendler1999causal}. Individuals are also challenged by their ability to make discrete decisions and have difficulty maintaining economic stability~\cite{shirani2011taking}.
As a means of overcoming such issues, individuals may turn to their social network connections, who in turn offer informational and emotional support~\cite{dechoudhury2017language}. The process of support-seeking and caregiving is reciprocal in that the more stressful an event is, the more support a person seeks, which returns greater support from others~\cite{collins2000safe}. The social support provided by one's network is known to reduce stress levels and obtain adequate resources, a concept known as the \textit{buffering hypothesis}~\cite{cohen1985stress}. The importance of social support in stressful life events and the buffering hypothesis has been extensively studied in several domains, including but not limited to health studies~\cite{mitchell2014stressors}, psychology~\cite{jackson1992specifying}, and family studies~\cite{szkody2019stress}.

The social support that one may receive or provide during shocks may not be identical. Not only the nature of the shock event but also factors such as social status~\cite{smith2012status} and gender~\cite{liebler2002gender} determine the type or impact of the support given. A distinct characteristic of social networks that we will focus on is the type of \textit{interpersonal relationship} between the support provider and receiver. It is known that a person's support system consists of various types of relationships such as spouses, immediate family, close friends, and social acquaintances~\cite{vaux1985support}, which can provide different types of support. Friends and neighbors within close proximity are capable of providing more instantaneous and instrumental support than family members, but families in turn provide more stable support that is unaffected by temporal factors~\cite{wellman1990different}. Work-related stress can be alleviated through support from friends and workplace relationships rather than other relationships~\cite{henderson1985social}. Overall, different relationships are capable of delivering different types and levels of support depending on the stress event.

\subsection{Shock responses in online social networks}
Researchers have increasingly been turning towards online social networks such as Twitter and Facebook for studying interactions during shock events. The formation and structure of online social networks mirror those formed offline~\cite{dunbar2015structure}, and the abundance, as well as the accessibility of interaction data among users has made it a popular research subject. People may post support-seeking messages on networking services visible to others to receive social support~\cite{oh2016impression}. As a response to the messages, one's neighboring users may choose to send messages of support, which can lead to increased levels of well-being~\cite{burke2016relationship}. Previous studies have examined the roles of users in online communities for providing support on cases such as medical issues~\cite{huh2016personas} or suicidal ideation~\cite{dechoudhury2017language}. In the context of unexpected life events, some studies have looked at changes in the structures of online social networks following events such as the death of a close friend~\cite{hobbs2017connective}, unemployment~\cite{gee2017social}, or breakups~\cite{garimella2014love}. While these studies examine events similar to those considered in our work, they do not provide comparisons of behavioral differences among relationship types.

\subsection{Research Questions}
\label{sec:rq}
Our research identifies shock-response behaviors that occur in online social networks, with a focus on the interpersonal relationships among the users. Given different social expectations for certain relationships (e.g., friends vs. co-workers), we hypothesize that relationship type provides different degrees of support following shock events and that this behavior which has been observed in offline studies can also be found in online social networks, even if the type of support and relationships who provide it vary from offline to online. Therefore, we formulate our first research question to investigate whether the activation levels in response to a shock differ by shock, and more importantly, whether there exist differences among relationships. 
\\
\textbf{RQ 1} \textit{Does the likelihood to respond to a certain type of shock differ by relationship type?}
\\

The closeness between two individuals is a well-known factor for determining whether support will be provided upon experiencing a shock~\cite{collins2000safe}. However, it is not well understood whether closeness is a determinant of support across different types of shocks and relationships. In addition, closeness can be measured in different ways, such as through communication frequency or structural embeddedness in a network. While these two measures are correlated, they measure fundamentally different aspects of closeness, and it is possible to experience cases such as ties that have high communication frequency but without any common connections~\cite{park2018strength}. Thus, we investigate the role of closeness, both tie strength and embeddedness, in the likelihood of response across relationship and shock types.
\\
\textbf{RQ 2} \textit{Does the degree to which tie strength and structural embeddedness affect responsiveness to shocks differ by relationship type?}
\\

Finally, we examine the shift in the context of communication between a dyad in response to a shock. Online social networks are both platforms for obtaining and sharing information and also for maintaining contact with existing social ties~\cite{kwak2010twitter}. While a user's social network is thus expected to contain messages of both social and informational content, exposure to a shock may rapidly shift the composition of the topics surrounding her. Specifically, the social buffering hypothesis suggests neighboring users will offer some form of support to the shocked user which comes at the expense of their original roles of sharing other types of information. We expect that during a shock, neighboring users will modify the topics of the messages that they share with the shocked user to provide support. Our third research question investigates how this shift occurs differently for each relationship and for each shock.
\\
\textbf{RQ 3} \textit{What are the topics that increase or decrease for each relationship type following a shock?}

\section{Dataset}
\label{sec:data}
To study how social behavior varies after shock events, we create and introduce a new dataset of Twitter users undergoing specific types of personal events and the social interactions they have after. Following, we describe how the shock events are identified and how the dataset is created to enable a pseudo-causal analysis of the effects of a shock.

\subsection{Identifying Shock Events}
\label{sec:id_shock}
Since our study focuses on the changes in a person's online social network that can be caused by exposure to shock events, we introduce a new data collection procedure to capture a large and accurate sample of shock instances and the network activity around the time of the shock. We begin by identifying four types of well-studied life events as shocks:

\begin{enumerate}
    \item\textbf{Romantic breakups} Romantic relationships provide a strong source of attachment that deepens over time and leads to strong levels of intimacy~\cite{hartup99romanticrelationships}. Ending a relationship through a breakup can lead to heightened levels of depression and anxiety~\cite{sprecher1998factors}. As a replacement, individuals undergoing breakups may turn towards other close members of their social networks, friends, and family members to compensate for the lost relationship~\cite{moller2003relationship}.
    \item\textbf{Exposure to crime} Being the victim of a crime is an uncontrollable life event that can cause negative effects on one's emotional, physical, and financial status~\cite{cutrona1990type}. Social support from others can help the shocked individual mitigate negative psychological effects caused by the incident and obtain guidance so that they can go through appropriate measures and solve potential issues~\cite{mason1996effect}.
    \item \textbf{Death of a close person}
    The death of a close person such as a friend or family member is, without doubt, a very stressful event that leads to strong negative emotions such as loneliness, and depression~\cite{burton2006bereavement}. Social support from others can help reduce such levels of loneliness, and people close to the bereaved can form new connections as a means of coping with grief~\cite{hobbs2017connective}.
    \item\textbf{Unexpected job loss}
    Unemployment can lead to a decline in one's time structure, social contacts, and activity levels~\cite{jahoda1981work}. The impact of job loss increases depending on financial situations and attachment to the job~\cite{leanai1990individual} and the loss of friends and colleagues, harming one's social support network~\cite{morris1992unemployment}. Social networks can enhance an individual's mobility by providing information and various resources~\cite{podolny1997resources}. 
\end{enumerate}
While all four categories of shocks are known to cause high levels of distress and thus have the afflicted call for social support, the social groups or relationships that would respond may actually differ by shock type. For instance, a colleague from work would respond differently to one's breakup versus one's sudden unemployment and choose to provide different levels of support. Additionally, being inflicted by these shock events is known to cause disruptions in one's online social network connections~\citep{garimella2014love,hobbs2017connective,burke2013using,deal2020definitely}. We can thus expect to identify varying degrees of interactions by relationship type, which differs by shock type.

Our approach, described next, involves a series of active learning-based filtering that combines regular expressions and deep learning text classifiers to extract a dataset of tweets describing a shock event, denoted as \textit{shock tweets}.

To be considered as a shock tweet, we set a number of requirements that each tweet needs to satisfy:
 \begin{enumerate}
    \item \textbf{Topic relevance}: The event described in each tweet should fall into one of the categories defined earlier as shock events: \textit{romantic breakups}, \textit{crime}, \textit{death of a close person}, or \textit{unexpected job loss}. 
    \item \textbf{Recency}: The event described in each tweet should be about an event that has happened ``recently". We preserve tweets that contain phrases describing that this event happened ``recently''~(refer to Table~\ref{tab:regex} in the Appendix) or within a week. This helps remove tweets posting shock events that happened in the past and have already been resolved.
    \item \textbf{Self-centeredness}: The event described in each tweet should be about an event that happened to the author. This prevents capturing tweets about events that happened to others, which we do not consider as personal life events. An exception is the case of shocks from the `death' category, where we only use tweets that describe the death events of a close person (e.g., a friend or family member).
    \item \textbf{General tweets}: We limit our scope to \textit{general tweets} or tweets addressed to all followers of a user account\footnote{https://help.twitter.com/en/using-twitter/types-of-tweets}. This differs from replies, which although publicly visible, are more targeted towards the original tweet's author rather than the entire followers of a user and will thus be processed differently by others. We also remove all retweets or quotes as we are only interested in the messages directly generated by the shocked user.
\end{enumerate}
We construct a set of regular expressions to match these requirements. For each shock, we list all possible ways to address a shock event (e.g., ``passed away'', ``died'', ``passing of''), a recent event (e.g., ``last week'', ``yesterday'', ``this morning''), and a close person in the case of death shocks.
Our source data is comprised of a 10\% percent sample of tweets produced between January 2019 and June 2020 obtained through the Twitter Decahose API. We remove reply tweets, retweets, and replies with comments to another tweet. We then apply regular expressions to filter out tweets that do not satisfy our conditions on topic relevance, recency, and self-centeredness~(Table~\ref{tab:regex} in the Appendix), resulting in an initial set of 178,416 candidate shock tweets~(Table~\ref{tab:num_tweets}).

\begin{table}[t]
  \centering
  \caption{The number of tweets retained after each step.}
\resizebox{0.45\textwidth}{!}{
\begin{tabular}{p{0.16\columnwidth}p{0.23\columnwidth}p{0.23\columnwidth}p{0.23\columnwidth}}

    Shock type & Regex filtering & Active learning & Covariate matching\\
    \hline
    Breakup & 18,277 & 3,249 & 1,191 \\
    Crime & 83,246 & 4,743 & 1,762 \\
    Death & 69,707 & 22,870 & 9,456\\
    Job loss & 7,186 & 2,171 & 1,160 \\
    \hline
    Total & 178,416 & 33,033 & 13,569 \\
  \label{tab:num_tweets}
\end{tabular}}
\end{table}

\begin{table}[h]
  \small
  \centering
  \caption{{The precision of valid shock tweets after filtering on regular expressions only.}}
  \resizebox{0.4\textwidth}{!}{\begin{tabular}{p{0.23\columnwidth}p{0.55\columnwidth}}
    Shock type & Precision of regular expressions \\
    \hline
    Breakup & 0.147 \\
    Crime & 0.048 \\
    Death & 0.218 \\
    Job loss & 0.202 \\
  \label{tab:regex_fraction}
\end{tabular}}
\end{table}

Even after applying the regular expressions, we observe that the precision of the filtered tweets is low~(Table~\ref{tab:regex_fraction}). As a next step, we improve the precision of our dataset by providing labels for the tweets and training a text classifier model through manual annotation. To ensure annotation quality, we randomly select 50 tweets per shock type and have three annotators with sufficient background knowledge to determine whether each tweet is a shock tweet or not. We then measure the Krippendorff's $\alpha$~\cite{hayes2007krippendorff} to measure inter-annotator agreement, where we achieve high agreement scores of 0.875 (breakup), 0.894 (crime), 0.891 (death) and 0.869 (job loss).
Once high levels of agreement are ensured, we then train multiple rounds of text classifiers accompanied by augmenting annotated samples for each round, which is a form of active learning~\cite{settles2009active}.
For each shock type an annotator provides labels for 1,000 tweets from our regex-filtered set of tweets, which we divide into a train/test/validation split of 400/400/200. This dataset is then used to train a classifier that predicts whether a given tweet is a shock tweet for that respective shock type. We use the pre-trained cased BERTweet~\cite{nguyen2020bertweet} model in Pytorch 1.8. For each model, we train for 50 epochs and save the model with the highest F1 score on the validation set, which is used to compute the F1 score on the test set. We use Adam~\cite{kingma2015adam} with a learning rate of 1e-8 and 100 warmup steps. 

\begin{figure}
    \centering
    \includegraphics[width=0.99\columnwidth]{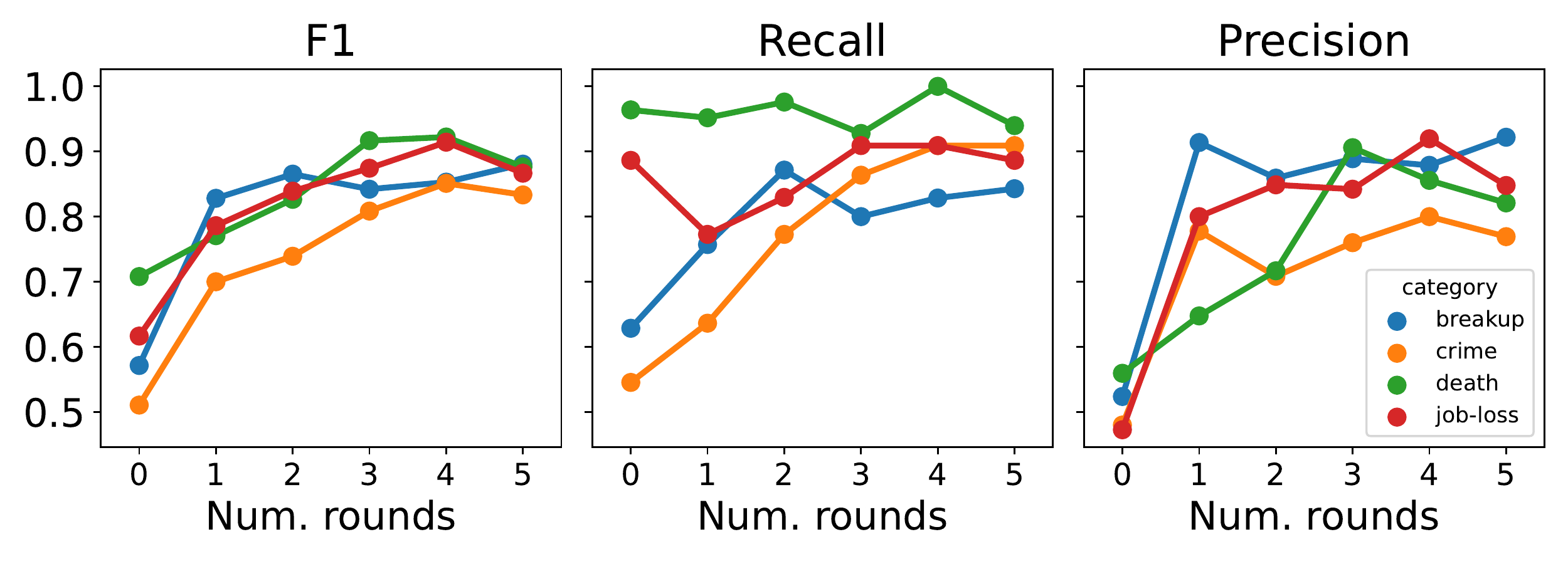}
    \caption{Comparison of metrics on the test set after performing additional rounds of active learning.}
    \label{fig:classifier}
\end{figure}

Once our initial classifier is trained, we improve performance through rounds of further labeling and training with additional samples. We use the initially trained classifier to infer the probability values of being a shock tweet, a continuous score ranging between 0 and 1, for all unlabeled tweets. We then select 200 tweets of which the inferred score is closest to 0.5, the decision boundary, and additionally annotate these selected tweets, which are added to the training set to train a new classifier model using identical hyperparameter settings. We repeat this process for five rounds, after which the performance  plateaus~(Figure~\ref{fig:classifier}). We then select the best-performing classifiers for each shock type, which produce F1 scores of 0.88 (breakup), 0.85 (crime), 0.92 (death), and 0.91 (job loss). Using these models we infer the remaining unlabeled tweets and preserve the samples whose inferred score is higher than 0.5. We combine these samples with the positively labeled samples to use as our set of 33,033 shock tweets. The number of tweets preserved through each step is shown in Table~\ref{tab:num_tweets}.

\subsection{Propensity Score Matching}
\label{sec:psm}
In order to identify what causal effects the shock event has on social ties' behaviors, we need to control for counterfactual settings of users who would have experienced a shock but did not, which exist in our initial observational data. For each shock tweet, we assume that the author of the tweet is a treated user, and so for users who posted multiple shock tweets (0.39\% of the treated users), we only considered the earliest instance as a valid shock tweet.
By comparing our treated users against a control group with similar covariates, we can ensure robustness in measuring the effect of self-disclosing a shock tweet and removing potential confounding effects. Therefore, we adopt a causal inference approach and use propensity score matching (PSM) to create a control set of users who have similar covariates as those undergoing a shock event~\cite{imbens2015causal}. 
As a matched user may change one's behavior over time, we match control users experiencing a shock with another user according to their behavior on a specific date.

\subsubsection{Selecting covariates for matching}
The first step of PSM requires identifying adequate covariates for each user to be later used in the matching task. These covariates should be relevant to whether a user posts a shock tweet or not or are likely to undergo the shock event itself. For every active user in our dataset, we consider the following covariates:
\begin{itemize}
    \item \textbf{User demographics}: As a proxy for actual demographic information of a Twitter user, we obtain inferred gender and age using \textsc{M3} model~\cite{wang2019m3inference}. The M3 model returns a distribution over age categories and a continuous score of gender performance\footnote{For age, we drop one category due to collinearity.}.
    \item \textbf{Twitter account properties}: We use the number of followers, friends (followers), and activities of each user as covariates. As these numbers change slightly over time, we use the earliest available measure in our dataset.
    \item \textbf{Social Condition}: We obtain inferred geolocations of Twitter users from the total variation geoinference method~\cite{compton2014geotagging} and the user's corresponding US census tract. We use the American Community Survey 2018 data for each tract to associate the user with the following covariates as proxies of the users' social conditions: racial distribution, ratio of income to poverty, education level, marriage rate, divorce rate, industry group composition, Gini index, unemployment ratio, and health insurance cover rate.
    \item \textbf{User activity levels}: To account for variation in the attention a user receives and their relative activity levels over time, we obtain the number of tweets (replies, mentions, retweets, and quotes) up to seven days before a particular date and aggregate them by tweet type.
\end{itemize}

For the initial set of control users, we consider all users active during the time range of our initial Twitter dataset (Jan. 2019 to Jun. 2020). We then preserve the users for which we can identify all covariates, which leaves 5,035,811 candidate users. We identify the same covariates for treated users, which reduces the number of valid treated users to 13,569~(Table~\ref{tab:num_tweets}).

\begin{figure}
    \centering
    \includegraphics[width=.96\columnwidth]{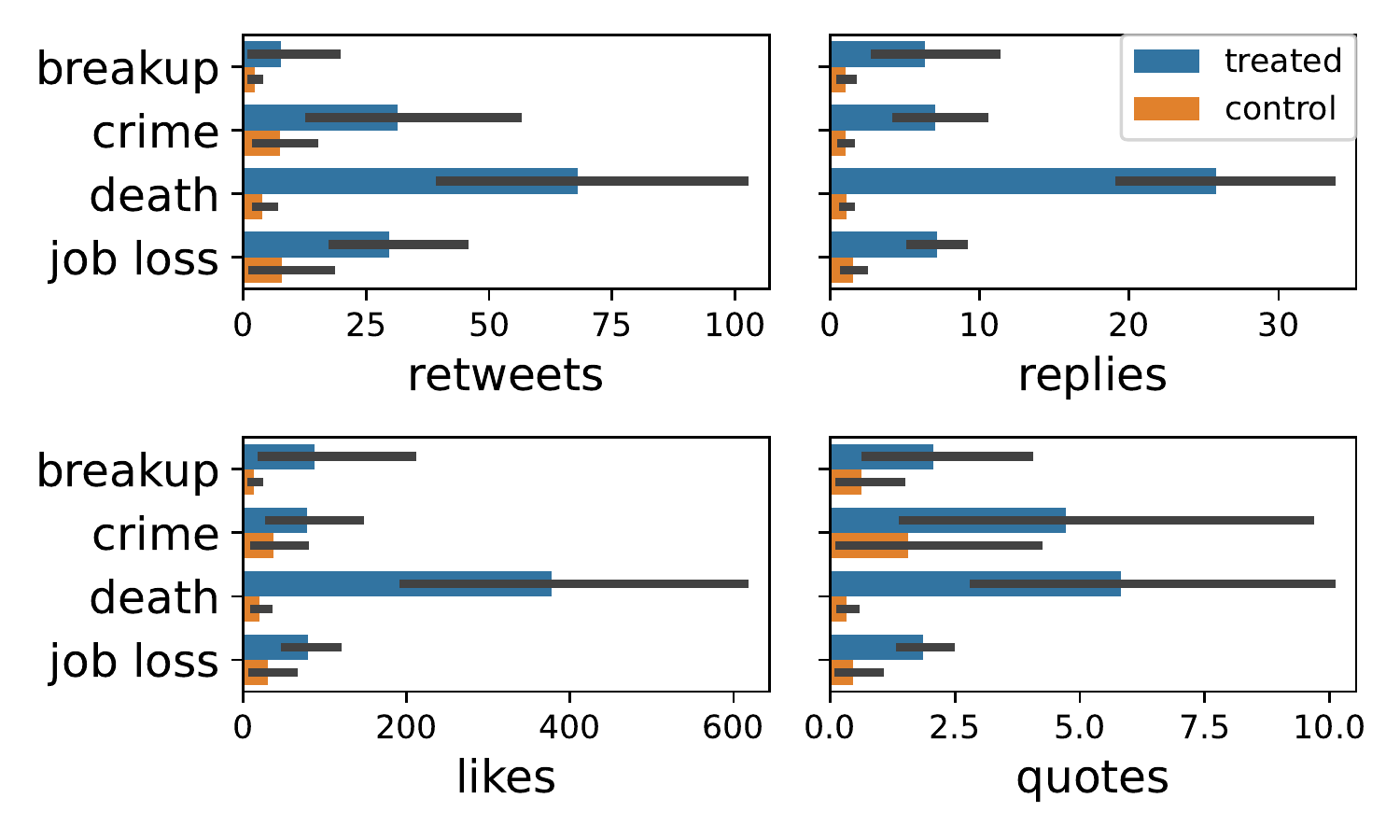}
    \caption{Comparison of response metrics between shock tweets and control tweets. Shock tweets receive a substantially larger amount of retweets, replies, likes, and quotes compared to the control tweets.}
    \label{fig:tweet_metrics}
\end{figure}

\begin{figure}
    \centering
    \includegraphics[width=0.86\columnwidth]{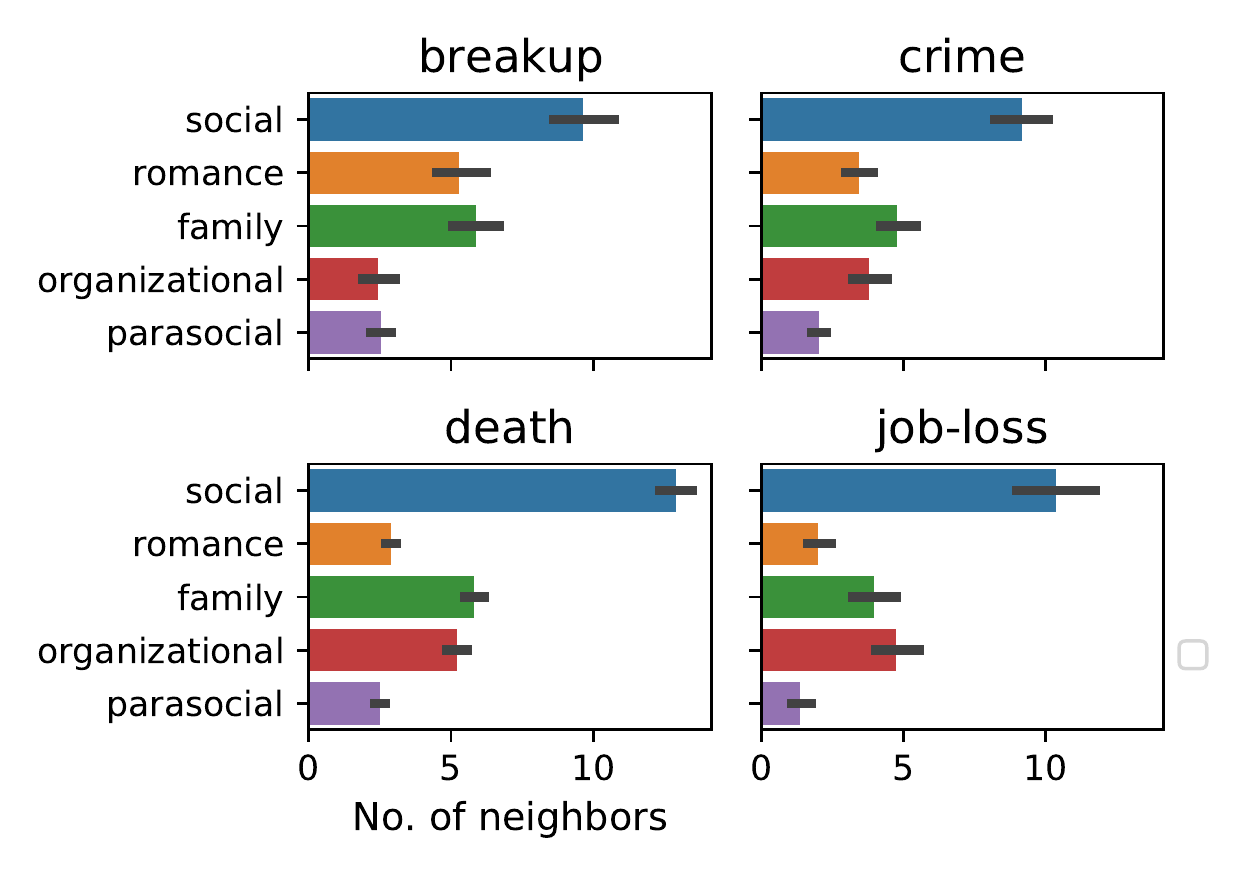}
    \caption{The average number of neighbors of each predicted relationship type for shock type.}
    \label{fig:rel_distribution}
\end{figure}

\subsubsection{Propensity Matching Users}
Instead of performing propensity matching at a user level, we perform matching at the user-date level. In other words, the same user exists through several different samples, as the user's activity covariates differ by each week. We include a matching candidate for a control user for their behavior each day they were active, referred to as a \textit{user-date pair};  this process allows closely matching the behavior of the treated user.  Following standard practices of PSM, we estimate propensities using a logistic regression model where the objective is to predict whether a user-date pair is a shock instance based on the associated user-date covariates~\cite{eckles2021bias}. The fitted model is then used to infer the propensity scores in [$0,1$] for every instance using the same covariates. PSM assumes the model learns to predict the likelihood of a shock from the provided covariates, and thus similar propensity scores will indicate similar covariates. Following \citet{eckles2021bias}, we sort the propensity scores for each user-date pair and divide them into $n$ strata of equal sizes, with the root of the number of treated samples as $n$. Note that $n$ is different for each shock type. Once both treated and control user-date pairs are sorted into the bins, we match each treated user with five randomly sampled control users whose behavior was recorded on the same day as the treated user, similar to the 1:5 ratio used in \citet{maldeniya2020herding}.

To verify the quality of our matching process, we measure the Cohen's \textit{d} effect size for each covariate by comparing the distribution obtained from the treated users with that of the matched users. For all covariates, we obtain an effect size of lesser than 0.269, which corresponds to the group status of shocked or matched user-date pairs accounting for lesser than 1\% of the variance~\cite{imbens2015causal}.

\subsection{Identifying relationships}
\label{sec:id_rel}
Social relationships can take on many distinct types, from general categories like family and friends to more specific types like step-parent or direct-report~\cite{wellman1990different}. Here, we adopt the model of \citet{choi2021more} that infers five relationship categories from Twitter interactions: \textit{social}, \textit{romance}, \textit{family}, \textit{organizational}, and \textit{parasocial}. These high-level categories capture the majority of relationship types expected to be observed on Twitter and, critically, have different social expectations in their behavior. 

\begin{figure*}
    \centering
    \includegraphics[width=1.9\columnwidth]{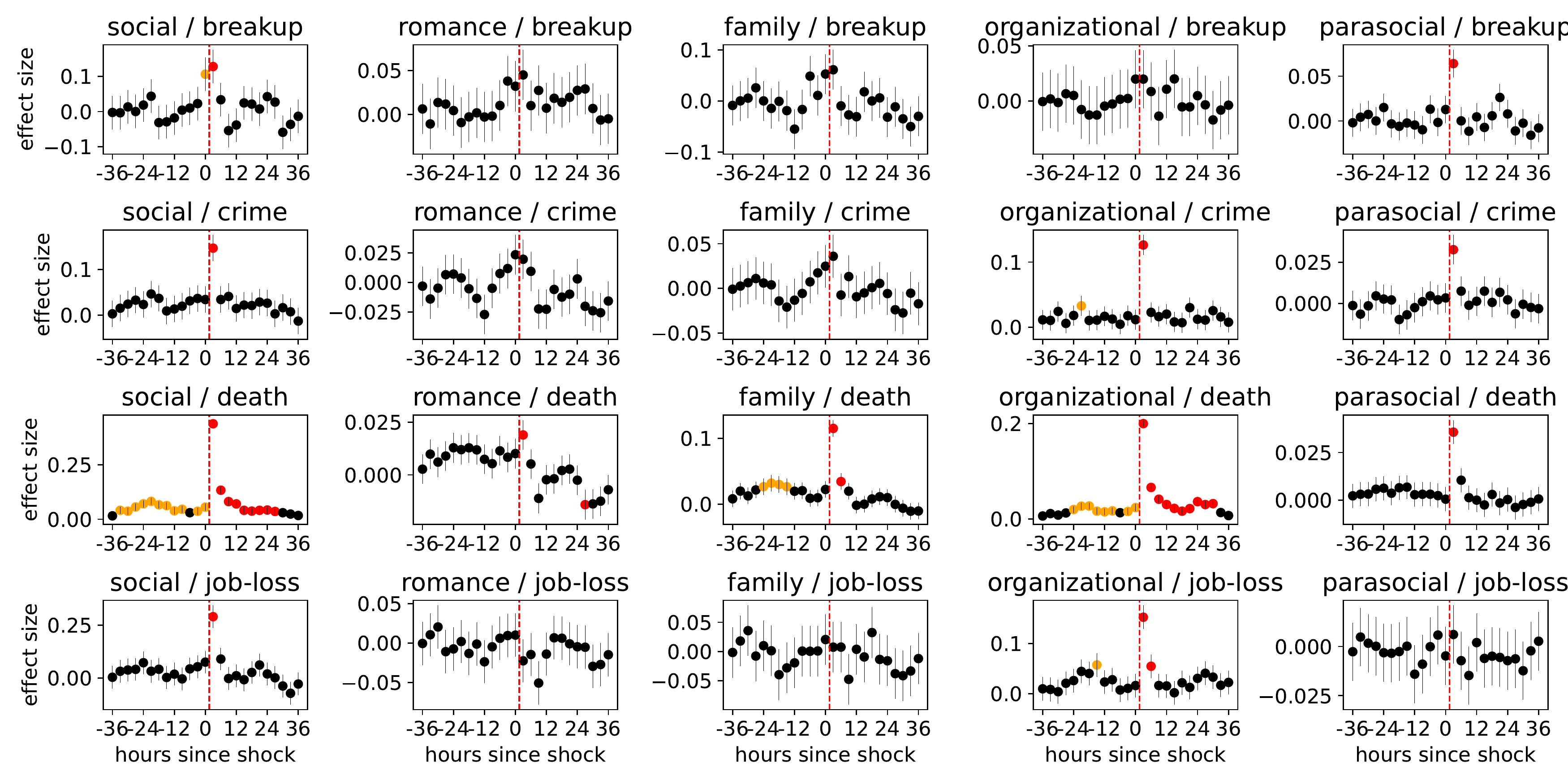}
    \caption{Changes in the volume of replies to shocked users across 3-hour blocks, measured using diff-in-diff relative to control users. The x-axis is the number of hours relative to the shock (e.g., the first hour of the shock corresponds to x=1). Statistically significant values (Bonferroni corrected) are colored in {\color{orange} orange} if before the shock and {\color{red}red} if after the shock. The red dashed line indicates the time of posting the shock tweet.}
    \label{fig:hourly_replied}
\end{figure*}

For the matched treated and control users, we identify all dyads that exchanged at least three interactions from the beginning of the time frame of our dataset to seven days before the shock event. Each tie is then classified using its text content, network features, and user description. Our selection to drop all interactions made after seven days prior to the shock tweet is to ensure that the relationship prediction is completely independent of interactions related to the shock itself. We compare our predictions with the results of another setting where we include the interactions around the time of the shock, and find that the predicted relationships of the two settings match  78\% (breakup), 80\% (crime), 83\% (death), and 86\% (job loss) of the time.

While the Choi et al. model requires at least 5 interactions, we relax the minimum amount of required interactions to three tweets as our time frame is much shorter (1.5 years vs. 7 years), leading to a smaller number of classifiable dyads. Only using tweets up to one week prior to the shock event avoids potential confounds from inferring relationships based on interactions during the shock event itself. To test the robustness of this setting, we identify dyads with 5 or more interactions, then create two feature sets of all dyads - one using only three interactions per dyad, and the other using all available interactions. We observe that the predicted relationship categories for the two settings match in 83\% of the predicted dyads, largely exceeding a random baseline of 20\% and showing that the predictions are robust in the number of interactions.

The distributions of the relationship types for treated and control users of each shock type are shown in Figure~\ref{fig:rel_distribution}. Across all shock types, the social relationship type is most common, consistent with \citet{choi2021more}. We also observe that the size of the average number of ties per relationship type is similar across shocks.

\section{Relationships and shock response}
\label{sec:diff}
The shocked users receive far more engagement from their peers~(Figure~\ref{fig:tweet_metrics}). Across all four types of shocks, the treated users received more retweets, replies, likes, and quotes of their tweets than those in the propensity-matched control users, suggesting that their social network is aware and responsive to these events while using their own networks to increase visibility and utilize social capital~\citep{boyd2010retweet}. Does this responsiveness vary based on the relationship the other user, referred to as an \textit{alter}, has with the shocked user and the nature of the shock? Here, we test RQ1 using a pseudo-causal difference-in-difference to test the effect of another person communicating with the person experiencing a shock.  

\subsection{Experiment setup}
\label{sec:experiment}
For each user-date pair, we collect all tweet activities involving the user within $\pm36$ hours of the timestamp of the corresponding shock tweet. This timeframe represents one full day (24 hours) before and after the event, plus 12 hours to account for time zone differences. We group the tweets into 3-hour bins, where we count the activities conditioning on (1) the direction (the shocked user responding to an alter user vs. an alter user responding to the shocked user), (2) the relationship type (five categories), and (3) the shock type (four categories). For each time series in the 80 conditions, we construct a difference-in-difference model with lead and lag variables. This model is summarized as follows:
\begin{displaymath}
  y_{d,t,s}=\alpha+D_d+T_t+S_s+\pi_tP_t+\epsilon_t
\end{displaymath}
where $y_{d,t,s}$ indicates the number of replies a user in strata $s$ received treatment at day $d$ with $t$ hours since the posting of the shock tweet, divided by the number of neighbors of the corresponding relationship type. $d$ is the date of the tweet (e.g., 2019-01-05) represented as a category variable, $t$ is a categorical variable for 3-hour intervals from -36 to +36, and $s$ is the corresponding strata of the user-date pair which was used for stratifying based on propensity scores. $D_d$ and $T_t$ capture temporal trends while $S_s$ captures differences between strata. $P$ is the treatment interaction of the relative hour from a shock tweet, which allows $\pi$ to capture not only post-treatment effects but also any pre-treatment effects around the shock. The effect sizes correspond to changes in the attention from the shocked user's neighbors.

\subsection{Results}
Social ties strongly engaged with individuals experiencing a shock but engagement varied significantly by the type of relationship and shock (Figure~\ref{fig:hourly_replied}). We highlight three findings.

\textbf{Romance and family ties are not as responsive online}
Studies on stress management through social interactions focus on the role of ``close'' relationships which consist of family members or romance partners~\cite{collins2000safe}. In contrast, we observe that in online social settings, the largest effect sizes on response rates come from social and organizational relationships. Indeed, the effect sizes of romance and family relationships are insignificant for all shocks apart from death, which is also much lower than those of social and organizational ties. Interestingly, replies from individuals with romantic ties \textit{drop} shortly after experiencing a death shock, an effect not observed in any other case. One possible explanation is that, unlike social or organizational ties where online social networks serve as the main channel for communication, romance or family ties would be able to communicate through other offline means such as phone calls, text messages, or in-person conversations~\cite{burke2014growing}; the gravity of a death shock could potentially decrease the alter's online communication, leading to the observed drop.

\textbf{Responses to shocks can be relationship-specific}
Based on the nature of the shock, individuals in certain relationships were more likely to respond---or not respond at all. This trend can be observed in the case of organizational ties, where in contrast to its high responsiveness for crime, death, and job loss shocks, individuals with organizational relationships had no significant increase in responding to breakup shocks. Similarly, parasocial relationships show significant levels of responsiveness in all shock categories except for job loss. These trends likely reflect normative boundaries in which content people share, which differ by relationship type; as a result, alters may feel uncomfortable showing interest in events that are considered out of their social boundaries~\cite{collins2000safe}.

\textbf{Pre-shock effects for death shocks}
For death shocks, a small but significant pre-shock effect can be seen before the actual shock. By manually examining a portion of the pre-shock activity, we were able to discover that this effect was partly due to announcements that their close ones were in a critical state,  such as someone getting injured and being transported to the ER or entering a critical condition, which are stressful situations themselves. These foreboding tweets can also be seen as request signals for support~\cite{oh2016impression} along with the shock tweet, 
which is returned by increased attention and support from others. 

\section{The impact of closeness on shock response}
\label{sec:closeness}
While our findings showed relationship-specific trends that differ for each shock type, these properties may differ even  for the same relationship, depending on other network properties of the dyad such as tie strength and embeddedness. Here, we formulate a task of predicting whether an interaction occurs within a dyad when one user posts a shock tweet, and investigate how tie strength and embeddedness play different roles across relationship and shock types.

\subsection{Experiment setup}
We examine how responsiveness can be predicted with respect to tie strength and structural embeddedness, two well-known proxies of closeness in a social network. We construct an undirected network using 10\% of all Twitter mention activities during a time frame of three months before the shock occurrence, where an edge is formed if a user mentions the other. We compute (1) tie strength, measured as the number of mentions an alter made to a shock user during that period, and (2) structural embeddedness, measured through Jaccard similarity which is the size of mutual neighbors of two users divided by the size of the union of their neighbors.
For a dyad of the shocked user $i$ and alter $j$ with a known relationship, we assign a label of `responded' (1) or `not responded' (0) to $y_{i,j}$ based on whether the alter user responded to the shocked user within 36 hours of the shock tweet.
We then formulate a logistic regression task with the following formula:
\begin{displaymath}
  y_{i,j}=\alpha+t_{i,j}+s_{i,j}+X_i+X_j
\end{displaymath}
where the number of mentions towards the shocked user $t_{i,j}$ represents tie strength and the Jaccard similarity $s_{i,j}$ represents structural embeddedness. The number of followers, friends and posts, the in-degree and out-degree of each user (vectors $X_i$ and $X_j$) are also included as independent variables which we control for.

\begin{figure}
    \centering
    \includegraphics[width=0.9\columnwidth]{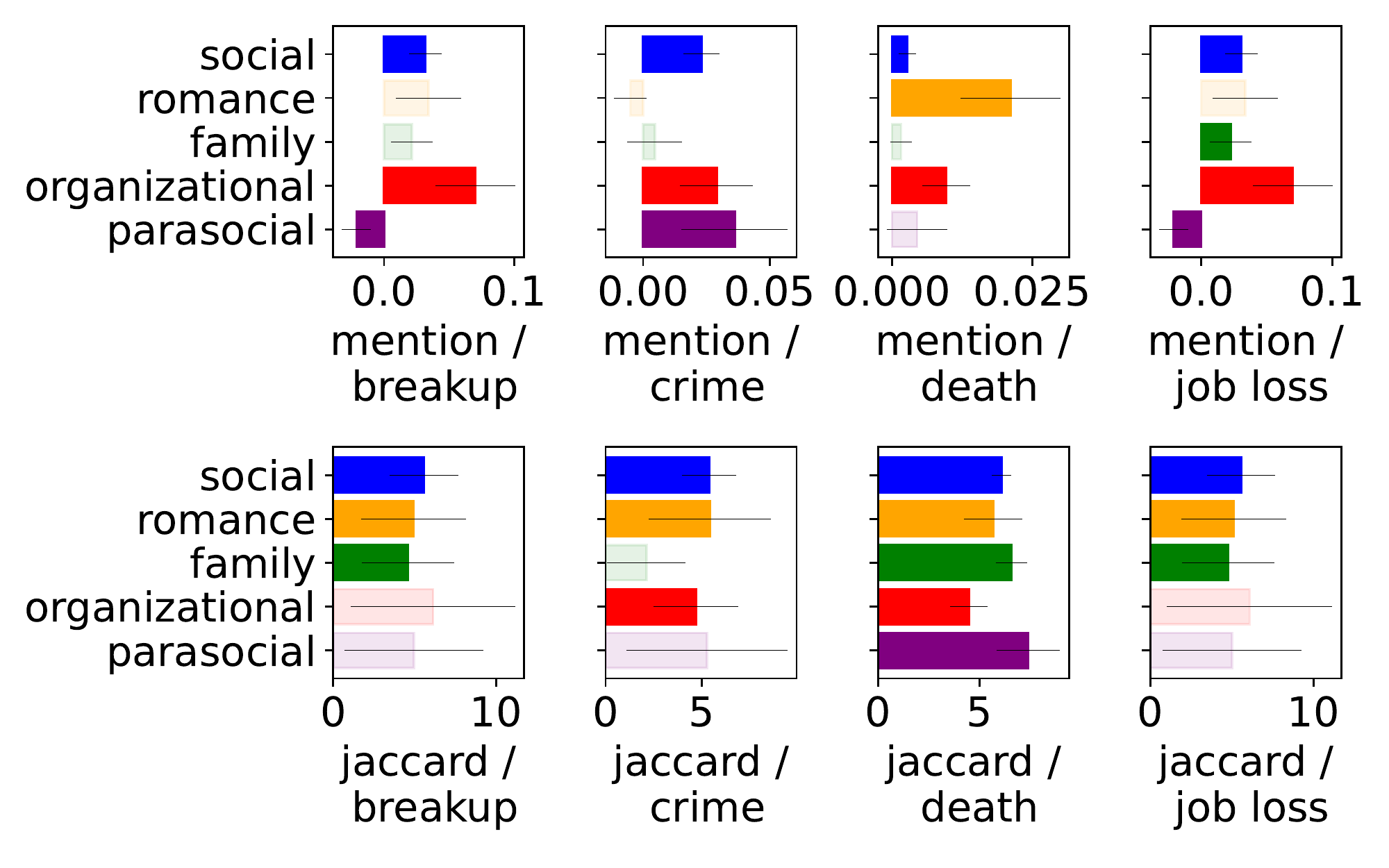}
    \caption{Coefficient sizes obtained from a logistic regression of response on mention frequency (tie strength) and Jaccard similarity (structural embeddedness). Values with solid colors indicate statistically significant (Bonferroni-corrected) coefficient values.}
    \label{fig:monthly_regression}
\end{figure}

\subsection{Results}
We display the coefficients of the network properties for each relationship-shock pair and compare them across relationship types for each shock in Figure~\ref{fig:monthly_regression}.
When significant, close ties are almost always likely to have a positive effect on response prediction, which is expected given the relationship between closeness and social support. Comparing relationship and shock type provides us with a clearer picture.
For example, in family relationships, mention frequency is insignificant for most shocks while the Jaccard similarity is almost always significant. This shows that social support from family members depends more on social embeddedness rather than communication frequency. The situation reverses for organizational ties, where communication frequency is more important than embeddedness across shocks. In addition, response during death events depends much more on embeddedness than the frequency of communication.
In summary, we generally confirm the important role of closeness in the likelihood that a social connection provides support during shocks, but at the same time, we find it is highly dependent on the type of closeness, shock, and relationship.

\section{Topic shifts in shock responses}
\label{sec:topic}

\begin{table}[t]
    \small
    \centering
    \caption{Top keywords and annotated topics from resulting topic model}
    \resizebox{0.42\textwidth}{!}{
    \begin{tabular}{p{0.04\columnwidth}p{0.78\columnwidth}p{0.20\columnwidth}}
    \toprule
    Topic no. & Keywords & Annotated topic \\
    \midrule
    1 & hair wear wearing black blue clothes red white & appearance \\
    2 & head face back hand hands eyes bit room & body \\
    3 & family loss love hear prayers sending condolences time & condolence \\
    4 & people it's don't that's i'm make good thing & conversations \\
    5 & it's i'm don't people good that's time thing & conversations \\
    6 & day great happy hope good birthday enjoy weekend & days \\
    7 & hope glad i'm man hear stay safe you're & emotional support \\
    8 & child mom baby kids age mother dad children & family \\
    9 & money work buy pay make free time stuff & finance \\
    10 & food eat chicken cheese eating pizza bread meat & food \\
    11 & drink water ice tea milk coffee cream bottle & food \\
    12 & game play games playing played switch it's xbox & games \\
    13 & balls demon harder image hill silent core sec & games \\
    14 & moon pokemon shiny sun trade frame giant sword & games \\
    15 & brain test order blood levels cancer care bubble & health \\
    16 & gay ship gender canon signs dark disaster ships & identity \\
    17 & lol shit i'm don't ass lmao fuck gonna & informal language \\
    18 & english language spanish speak words write german french & language \\
    19 & movie watch show good episode character season favorite & media \\
    20 & song music album listen songs love sing video & music \\
    21 & cats dogs pet yang ship they're tht died & pets \\
    22 & adopt home pledge foster rescue save give dog & pets \\
    23 & trump case court rule report issues legal cases & politics \\
    24 & black white women men people racist call support & social issues \\
    25 & trump vote party government state win american country & politics \\
    26 & love cute good amazing beautiful nice girl omg & positive emotions \\
    27 & god jesus soul church rest lord grant allah & religion \\
    28 & school high class year kids teacher learned college & school \\
    29 & twitter tweet post account lol send thought link & social media \\
    30 & red sox mlb fans white starts remains queue & sports \\
    31 & team year game win season play top players & sports \\
    32 & match champion wrestling wwe heel turn ring roman & sports \\
    33 & data apps desktop storage celica folders hai efficient & technology \\
    34 & time i’m back lol day years ago today & time \\
    35 & side city car big south back summer driving & traffic \\
	\bottomrule
	\end{tabular}
	}
    \label{tab:topics}
\end{table}

\begin{table*}[]
    \centering
    \small
    \resizebox{0.7\textwidth}{!}{
    \begin{tabular}{p{0.13\columnwidth}|p{0.26\columnwidth}p{0.26\columnwidth}p{0.26\columnwidth}p{0.26\columnwidth}p{0.26\columnwidth}}
        Shock & Social & Romance & Family & Organizational & Parasocial \\
        \hline
    Breakup 
        & \begin{tabular}[t]{@{}l@{}}\color{teal}conversations ↑\end{tabular}
        & 
        & 
        & 
        & \begin{tabular}[t]{@{}l@{}}\color{teal}informal language ↑\end{tabular} 
        \\ 
    Crime &
      \begin{tabular}[t]{@{}l@{}}\color{teal}condolence ↑\\ \color{teal}emotional support ↑\end{tabular} 
       & \begin{tabular}[t]{@{}l@{}}\\ \color{teal}emotional support ↑\end{tabular}
       & \begin{tabular}[t]{@{}l@{}}\\ \color{teal}emotional support ↑\end{tabular}
       & \begin{tabular}[t]{@{}l@{}}\\ \color{teal}emotional support ↑\end{tabular}
       & \begin{tabular}[t]{@{}l@{}}\\ \color{teal}emotional support ↑\end{tabular}
       \\ 
    Death &
      \begin{tabular}[t]{@{}l@{}}\color{red}appearance ↓\\ \color{red}body ↓\\ \color{teal}condolence ↑\\ \color{red}conversations ↓\\ \color{teal}emotional support ↑\\  \color{red}finance ↓\\ \color{red}food ↓\\ \color{red}games ↓\\ \color{red}informal language ↓\\ \color{red}media ↓\\ \color{red}music ↓\\ \color{red}pets ↓\\ \color{red}positive emotions ↓\\ \\ \color{red}social media ↓\\ \color{red}sports ↓\\ \color{red}time ↓\\ \color{red}traffic ↓\end{tabular} &
      \begin{tabular}[t]{@{}l@{}}\\ \\ \color{teal}condolence ↑\end{tabular} &
      \begin{tabular}[t]{@{}l@{}}\\ \\ \color{teal}condolence ↑\\ \\ \color{teal}emotional support ↑\\ \\ \\ \color{red}games ↓\\ \color{red}informal language ↓\\ \color{red}media ↓\\ \\ \color{teal}pets ↑\\ \\ \color{teal}religion ↑\\  \color{red}social media ↓\\\end{tabular} &
      \begin{tabular}[t]{@{}l@{}}\\ \color{red}body ↓\\ \color{teal}condolence ↑ \\ \color{red}conversations ↓\\ \color{teal}emotional support ↑\\   \color{red}finance ↓\\ \color{red}food ↓\\ \\ \color{red}informal language ↓\\ \color{red}media ↓\\ \\ \\ \\ \\ \\ \\ \color{red}time ↓\end{tabular} &
      \begin{tabular}[t]{@{}l@{}}\\ \\ \color{teal}condolence ↑\end{tabular} \\ 
    Job loss &
      \begin{tabular}[t]{@{}l@{}}\color{teal}condolence ↑\\ \color{teal}emotional support ↑\end{tabular}
      &
      &
      & \begin{tabular}[t]{@{}l@{}}\color{teal}condolence ↑\\ \color{teal}emotional support ↑\end{tabular} 
      &
    \\
    \end{tabular}}
    \caption{A comparison of topics that experienced significant increases or decreases following a shock. The arrow direction along with the color indicates whether the topic usage has {\color{teal}increased} or {\color{red}decreased} after a shock when compared with the control group. The numbers in brackets correspond to the topic numbers in Table~\ref{tab:topics}.}
    \label{tab:topics_by}
\end{table*}

While the response to another person's shock is likely to be of social support, the topics discussed take into account the information shared between the relationship and the underlying social background.
Our final question is whether users adjust their topics when responding to shocks and if this level of adjustment differs by relationship.

\subsection{Experiment setup}
By considering each tweet message as a document and running a probabilistic topic model~\cite{blei2003latent} across all documents, we can aggregate the output topic distributions for each document across shock type and relationship type. We can compare the topic distributions to interpret which relationship type is more likely to be associated with a particular topic. We aggregate the text content of the replies a treated or control user received during the first 36 hours since posting a shock tweet and run an LDA topic model MALLET~\cite{mcCallum2002MALLET} on 10, 20, 50, and 100 topics. Recent work~\citep{hoyle2021topicmodel} have shown that evaluation metrics such as coherence scores may not align well with human judgments on topic quality, and thus we manually examine the topic model results to select the topic size that contains the most diversity and least number of overlapping topics, leading to a size of 50.
%
As the topics produced by topic models are unsupervised, we manually label each topic after inspecting its most probable words; topic labels were obtained for 35 of the 50 topics; the remaining 15 did not have clear organization and were not used further in our study.
The most probable words of each topic and its label are shown in Table~\ref{tab:topics}. We observe a wide variety of topics, reflecting the range of information exchanged in Twitter conversations.

Once the topic distribution for each tweet is obtained, we examine whether posting a shock significantly increases or decreases the usage of specific topics seen in the replies from each relationship type. For each topic, relationship type, and shock type, we obtain two probability distributions from (1) the probability of that topic in replies toward shocked users and (2) the topic's probabilities in replies to control users. We then test whether the two distributions have an equal mean using the student's t-test. This results in a total of 1,000 comparison t-tests as there are 50 topics, 5 relationships, and 4 shocks. To account for the multiple comparisons, we apply Bonferroni correction by defining an effect as significant if and only if the p-value of the t-test is less than 0.05 divided by 1,000, imposing a strong threshold for significance. 

\subsection{Results}
We find 66 out of 1,000 possible cases across the combination of all shocks, relationships, and topics in which the change in usage is significant, shown in Table~\ref{tab:topics_by}. Following, we highlight three trends by shock type.

\textbf{Breakups} For breakup we observe two cases of significant increase in topic usage, coming from social (conversations; T4) and parasocial (informal language; T17) relationships. These topics are both situated in the `conversations' category which indicates an increased amount of casual conversation-based communication. When experiencing and posting a breakup, people considered in a social or parasocial relationship would likely engage more in conversations, either to discuss the event or to divert the attention away~\cite{moller2003relationship}. This shock type is also intriguing in that there is no apparent increase in condolence (T3) or emotional support (T7), which is observed in all other shocks. Support following breakups is shown not explicitly through words of condolence but instead in a form of increased interactions and conversations.

\textbf{Crime \& job loss} All relationship types responded to crime shocks by increasing emotional support (T7), and also condolence (T3) in social relationships. For job loss, we observe an increase in condolence and emotional support only from the social and organizational categories, consistent with our earlier findings from Section~\ref{sec:diff}.2.

\textbf{Death} Responses following a death shock show the most diverse changes in topic composition for all relationships. For romance and parasocial ties, we observe significant changes only in condolence (T3). Other relationships display more dynamic topic shifts. For social, family, and organizational relationships, along with the increase in condolence and emotional support (T7) we observe decreased usage of several topics ranging from appearance (appearance, body) to entertainment (games, media, sports). 
As for decreasing topics, we can notice that (1) relationships such as social ties discuss a wide variety of topics in pre-shock communication, and (2) in the case of death shocks these relationships are discouraged to transmit information other than messages of condolence and direct emotional support. We also find evidence of relationship-specific social support, seen in the case of topics related to pets (T22) and religion (T27) being increasingly used in family relationships.

In general, our results show that once a user experiences a shock, their neighbors may choose to significantly reduce the likelihood of conveying topics related to everyday events such as sports or politics, and shift towards expressing  condolences and support. At the same time, there are also relationship-specific topics that are exchanged, such as topics related to pets or religion from family members.

\section{Discussion}

\textbf{General discussion and implications}
One consistent theme throughout this study is that relationship types have their own functionalities, leading to different levels of engagement and communication following shocks. Our findings highlight the existence of latent social processes within the interactions of online social networks, which can be revealed only by explicitly modeling relationship types. Even well-established social dimensions such as tie strength and structural embeddedness which are known to contribute to responsiveness contribute differently by relationship. Our work suggests incorporating interpersonal relationships along with other network properties to gain a richer understanding of the dynamics within social networks.

Our study shows that existing theories on social relationships and social support do not always fit in the context of online social networks. Previous studies have emphasized the role of ``close'' relationships such as family members and romantic partners as major providers of social support. However, our results show that, in online settings, social and organizational ties are the most responsive and supportive, and also that closeness does not always lead to higher responsiveness. This discrepancy may result from several reasons, such as romance or family ties having other communicative means to provide support instead of Twitter~\cite{burke2014growing} or a disinclination to exchange personal conversations on public social networking sites.

Our work can help identify which ties in online networks are most supportive during shock experiences. For individuals experiencing stressful life events, it is often challenging to disclose their problems in public online spaces due to issues such as impression management concerns~\cite{oh2016impression}. The response behavior seen in our findings can help indicate the users in specific relationships that are likely to be receptive to outreach after different shock events.
Our findings can be used as a basis for providing topic- and relationship-specific support to shocked users by recommending the most suitable ties for different situations.

\textbf{Limitations}
Our work only examines shocks in an online setting, leaving open the question of how these trends hold in offline settings for the same events. Prior work has studied these offline behaviors separately~\cite{kendler1999causal}, yet the challenge of growing these studies to our scale is likely prohibitive. Future mixed-method work is needed to fruitfully identify how individuals leverage their offline \textit{and} online networks during shock events.

Similarly, we only examine public responses to life events. As evidenced by the large number of public shock event announcements and the many responses and expressions of support we found in our data, we believe this is an important phenomenon to study. We do acknowledge that considering private messages can lead to different findings. For instance, studying the pattern and content of private messages or how interactions differ from previous years where private messaging was unavailable could lead to interesting findings. However, we leave those to future studies.

Our study focused on short-term interactions following a shock. Although life shocks are often serious events, our results suggest the impact on one's online network was very brief, with the shocked user returning to normal activity within a week. However, the event may still lead to the long-term impact, such as the rewiring of network structure or changing ties, which longer-scale studies may identify.

Finally, the scope of our relationships is limited to five categories based on the classifier~\cite{choi2021more}. These categories may not capture the entirety of social relationships in online social networks. Further, while the best available model for our setting, the classifier itself may exhibit bias that affects our results. Future work on relationship inference could expand the scope of our analysis by identifying additional relationships or more specific subtypes of relationships to capture fine-grained trends.

\section{Conclusion}
In unexpected distressing circumstances, a helping hand or a kind word can make all the difference. Our results show that when individuals undergo unexpected shocks, others do reach out online but who specifically reaches out depends on the nature of their relationship and its interaction with the type of shock. Through a large pseudo-causal study of individuals experiencing four types of shocks paired with a precisely matched control set, we demonstrate that social relationships vary significantly in which types of shocks they engage with and even in the context of how they communicate to the shocked individual. Further, we show that while higher tie strength and social embeddedness both increase the likelihood of whether a person will respond in the case of most shocks, these trends are highly dependent on the relationship and shock event type.
The insights provide further evidence for the importance of explicitly modeling social relationships in studies on social network dynamics. We release our code, trained classifiers, and data at \emph{https://github.com/minjechoi/relationships-shocks}.

\section*{Ethical Statement}
We describe the possible ethical concerns of our study and our efforts to mitigate them. First, the data collection was performed while abiding by Twitter's Terms of Service on data collection and distribution,\footnote{https://developer.twitter.com/en/developer-terms}, where we only used tweets publicly available from the API at the time the study was conducted. 
Second, due to the sensitive nature of the events described by the users and Twitter's API terms of service, we will not make raw tweets publicly available. For reproducibility purposes, we will instead publish the Tweet IDs and our code. 
Third, we do not report findings that may compromise any individual's privacy. Our results represent the aggregated behavior of hundreds or thousands of users.
Fourth, our study is purely observational and we did not interfere, experiment, or interact with any user. Instead, the results of the paper were obtained through carefully designed quasi-causal methods. In summary, our findings were obtained and presented while minimizing privacy risks and ethical concerns through several measures.

\section*{Acknowledgments}
This work was supported by the National Science Foundation under Grant Nos. IIS-2007251 and IIS-2143529 and by the Air Force Office of Scientific Research under award number FA9550-19-1-0029. 

\appendix
\section{Appendix}

\begin{table}[h]
    \centering
    \small
    \caption{List of regular expressions used for initial filtering on shock tweets}
    \resizebox{0.45\textwidth}{!}{
    \begin{tabular}{p{0.84\columnwidth}p{0.16\columnwidth}}
    \toprule
    Regular expression & Filter \\
    \midrule
    \hline
    broken? up $|$ break ?up $|$ dumped $|$ divorce $|$ separated with $|$ end(?:ed)?(?: to)? (?:my$|$our$|$the$|$a)(?: \textbackslash~w+)\{0,2\} relationship $|$ parted ways & topic: breakup \\ \\
    robbed $|$ assault| \textbackslash~bstole $|$ mugged $|$ \textbackslash~brape $|$ discriminat $|$ offended $|$ harass $|$ \textbackslash~bstalk $|$ fraud $|$ lynch $|$ victim $|$ involved in a $|$ \textbackslash~babus(?:ed|ive) $|$ threaten $|$ broke  into $|$ \textbackslash~bhacked $|$ damaged $|$ \textbackslash~bsued\textbackslash~b $|$ \textbackslash~btowed $|$ accident\textbackslash~b $|$ stalk $|$ fraud $|$ arrest $|$ beat(?:en)? up & topic: crime \\ \\
    
    passed away $|$ death $|$ passing (?:of|away) $|$ died $|$ loss of my $|$ killed $|$ murder $|$ suicide $|$ lost (?:his|her|their) $|$ (?:life|lives) & topic: death \\ \\
           
    father $|$ dad $|$ mother $|$ mom $|$ mum $|$ gran\textbackslash~w+ $|$ gram\textbackslash~w+ $|$ aunt(?:\textbackslash~w+)? $|$ buddy $|$ pal $|$ uncle $|$ \textbackslash~bson $|$ daughter $|$ cousin $|$ niece $|$ nephew $|$ brother $|$ sister $|$ wife $|$ husband $|$ (?:boy|girl)friend $|$ \textbackslash~bfriend $|$ (?:\textbackslash~w+)?mate $|$ partner $|$ fianc(?:\textbackslash~w+) $|$ family $|$ relative & close:death \\ \\
    
    \textbackslash~bfired\textbackslash~b $|$ laid off $|$ furlough $|$ \textbackslash~blay ?off $|$ jobless $|$ unemployed $|$ lost (?:my|the)(?: \textbackslash~w+)\{,2\} (?:job|position) $|$ \textbackslash~blet go\textbackslash~b & topic: job loss \\ \\
                 
    today $|$ tonight $|$ yesterday $|$ \textbackslash~bjust\textbackslash~b $|$ recent $|$ \textbackslash~bnow\textbackslash~b |
            ( last $|$ this $|$ on $|$ at $|$ in the ) ( morning $|$ afternoon $|$ evening $|$ night $|$ monday $|$ tuesday $|$ wednesday $|$ thursday  $|$ 
            friday $|$ saturday $|$ sunday $|$ weekend $|$ week \textbackslash~b) $|$ ( days $|$ week $|$ (hour|minute)s?) (ago|since) & recent \\ \\
    (weeks $|$ month $|$ year $|$ yr) $|$ a while $|$ some time $|$ anniversar $|$ the day $|$ when \textbackslash~w+ was $|$ dream $|$ \textbackslash~b(in|at|around) [0-9]{1,4}\textbackslash~b & non-recent \\
    \end{tabular}
    }
    \label{tab:regex}
\end{table}


\bibliography{aaai22}

\end{document}